\documentstyle[a4,12pt]{article}

\input{epsf}

\begin{document}
\titlepage 
\begin{flushright}
CERN-TH/2000-111\\\
SPhT 00/46\\
hep-ph/0004133
\end{flushright}
\vskip 7pt
\begin{center}
{\bf \Large \bf
Models with Inverse Sfermion Mass Hierarchy and Decoupling of
the SUSY FCNC Effects}
\end{center}

\vskip 1.2cm
\begin{center}
{ \bf Ph. Brax\footnote{email: philippe.brax@cern.ch}\footnote{On leave from
 CEA-SACLAY, Service de Physique Th\'eorique, 
F-91191 Gif-sur-Yvette, France}}
\vskip 1pt
\end{center}
\centerline {Theory Division, CERN, CH-1211, Geneva, 23, Switzerland}
 
\vskip 1cm
\begin{center}
{ \bf C. A. Savoy\footnote{email:savoy@spht.saclay.cea.fr}}
\end{center}
\vskip 1pt
\centerline{ CEA-SACLAY, Service de Physique Th\'eorique, 
F-91191 Gif-sur-Yvette, France}
\vskip 1 cm
\leftline{Key Words: Beyond Standard Model, Supergravity Models}
\vskip 2cm

\abstract
{ We study the decoupling of the first two
squark and slepton families in order to lower the flavour changing
neutral current effects. Models with inverse sfermion mass hierarchy
based upon gauged $U(1)$ flavour symmetries provide a natural framework
where  decoupling can be implemented. Decoupling requires a large gap
between the Fermi scale and the supersymmetry breaking scale.
Maintaining the electroweak symmetry breaking at the Fermi scale
requires some fine-tuning that we investigate by solving the two-loop
renormalization group equations. We show that the two-loop effects are
governed by the anomaly compensated by the Green-Schwarz mechanism and
 can be determined from the quark and lepton masses. The
electroweak breaking constraints lead to a small $\mu$ scenario where
the LSP is Higgsino-like. }

\newpage

\section{Introduction}
Many sfermion mass patterns have been suggested in order to
 alleviate the effects of the flavour non-diagonal contributions
in the supersymmetric Flavour Changing Neutral Current (FCNC) problems.
Three main mechanisms have been proposed:\.  (i) {\it degeneracy} in
the scalar masses;\. (ii) {\it alignment} between fermion and sfermion
mass matrices;\. (iii) {\it decoupling} of all virtual supersymmetric
effects by large scalar masses.  Degeneracy  seems  natural  in the
context of gravity mediated sypersymmetry breaking although it
could also be pointed out that a generic flavour dependence of the
K\"{a}hler potential tends to spoil  this degeneracy in the presence of
a spontaneously broken flavour symmetry, if unprotected by this
symmetry.  In the framework of gauged flavour symmetries the induced
$D$-type soft masses are especially dangerous in that respect.
Actually, gauge mediation of supersymmetry breaking at a relatively low
scale would lead to negligible FCNC effects if the gravitino is light
enough but it apparently faces inherent problems to induce the correct
electroweak symmetry breaking.

The implementation of the alignment prescription via a rather ad hoc
choice of flavour symmetries has been advocated \cite{nls} to be  the
only natural solution, although some of the arguments used have to be
revised in the framework of broken supergravity theories. The examples
of such symmetries are sufficiently contrived that one feels compelled
to look for another possibility. On the other hand, decoupling has
always served as a possible remedy  for the inconsistencies of any
model beyond the standard theory.  However, third generation scalars
together with the Higgs and gaugino sectors which control the
electroweak symmetry breaking should be kept light enough to avoid too
much fine-tuning.  Since FCNC effects are more stringent within the
first two generations of quarks and leptons, it has been envisaged
that the first two generations could be considerably heavier than the
other supersymmetric particles\cite{dg} - actually, an efficient
suppression of FCNC effects by decoupling alone would require very
heavy masses.  As a matter of fact, the first and second families of
sfermion masses do enter the MSSM expression for the $Z$ boson mass
when one takes into account the  two loop effects . This has been used
to put limits on the mass gap between generations of scalars, mostly by
requiring the absence of an excessive fine-tuning.  Other bounds have
been obtained from the positivity of the stop masses\cite{mura}.  We
will comment on these generic bounds later on.

Our main interest is to cast the decoupling approach within
spontaneously broken supersymmetry scenarios.  In this context a very
large gap between families seems to be difficult to reach in the
supergravity mediated framework due to the form of the scalar masses,
assumed to be flavour diagonal, for simplicity,
\begin{equation} 
m_i^2= m_{3/2}^2 +\frac{\left< F_{a}\right> \left<
\bar F_{\bar b}\right> }{M_{P}^2}\partial_a\partial_{\bar b} \ln
(\partial _i \partial_{\bar i} K) +X_i \left< D_X\right>\ ,\label{smass}
\end{equation} 
where $K$ is the K\"{a}hler potential, $X_i$ is the gauge charge of the
state whilst $\left< F_a\right>$ and $\left< D_X\right>$ are the
auxiliary fields responsible for supersymmetry breaking, which
(vectorially) add up to $\sqrt{3} m_{3/2} M_P$.  This formula displays
the dependence of the sfermion masses on :({\sl i}) flavour independent
terms such as $m_{3/2}^2$;({\sl ii}) $\left< F_a\right>$ along
directions where there is a dependence of the K\"{a}hler metric in the
matter field families; ({\sl iii}) $\left< D_X\right>$ along flavour
symmetries.  One easily realizes from (\ref{smass}) that scalars in
different generations cannot be split by more than one order of
magnitude, unless ({\sl i}) is compensated by ({\sl ii}), like in
no-scale supergravity.

In the presence of supersymmetry breaking and flavour symmetry breaking
there are two sources of ``induced'' supersymmetry breaking capable to
yield flavour dependent scalar mass splittings. Let us consider a
simple situation\cite{FN} in order to illustrate this point: an Abelian
flavour symmetry $U(1)_X$ broken by the value of a Frogatt-Nielsen
field $\phi$ with $m_{3/2}\le \left<\phi\right> \le M_P$. Assuming a
vanishing cosmological constant, from the supergravity Lagrangians one
obtains\cite{dps,dd} the induced supersymmetry breaking $\left<
F_{\phi}\right> \sim m_{3/2}\left<\phi\right>$ along the $\phi$
direction and $\left< D_X\right>\sim m_{\phi}^2$ where $m_{3/2}^2$ is
the total supersymmetry breaking and $m_{\phi}^2$ is the $\phi$ soft
mass, of $O(m_{3/2}^2).$ In this case, the $D$-type splitting is always
relevant, the $F$-type one being proportional to $\left<\phi\right> ^2
/ M_P^2 .$
\subsection{\sl iMSSM}
Despite the strong motivations for models with very heavy scalars the
only concrete ones discussed so far are those coined "inverse hierarchy
models" (iMSSM). They are based on the assumption that an anomalous
$U(1)_X$ gauge symmetry is present  in the flavour sector of the
theory. The anomaly is fixed by the Green-Schwarz mechanism which then
determines the scale of the flavour symmetry breaking
$\left<\phi\right>.$ The induced $D$-term produces a mass splitting,
$m_i^2-m_j^2 = (X_i- X_j)m_{\phi}^2 .$ In models where the anomalous
$U(1)_X$ is responsible for the fermion mass hierarchy the charge
difference are roughly related to the fermions masses\cite{dps}
\begin{equation}
m_i^2-m_j^2 \propto \ln \left(\frac{m^F_j}{m^F_i}\right)\ .\label{hier}
\end{equation}
This leads to an inverse hierarchy in the sfermion masses compared to
the fermionic hierarchy (quarks and leptons).

This fact was first pointed out in the framework of general broken
supergravity coupled to Abelian flavour gauge symmetry\cite{dps} and,
subsequently, in models with dynamical supersymmetry breaking\cite{dd}.
It has been noticed that, because the top Yukawa coupling is of $O(1)$,
there is the relation $(X_{t_L}+X_{t_R}+X_{H_2})=0$, for the charges of
the top-Higgs sector, implying that the soft mass combination,
$(m_{U_3}^2+m_{Q_3}^2+m_{H_2}^2)$ receives no contribution from the
$D-$terms. Since this is the combination appearing in the one-loop
correction to the boson mass $M^2_Z$, the radiative gauge symmetry
breaking is automatically protected, at one-loop, against large
$D-$terms that are due to {\sl any} gauge symmetry broken at scales
below $M_P$.

It goes without saying that the FCNC effects are particularly dangerous in
the iMSSM framework. It has been suggested that this problem can be
evaded by a suitable combination of degeneracy, alignment and, last but
not least, decoupling of the first two generations\cite{dps,nel}. The
latter has prompted us to evaluate the two-loop corrections. This is
done in Section 2.
\subsection{\sl A limit on two-loop effects} 
In Section 3, we show that for inverse
hierarchy models based on $D$-term splitting the dominant contribution
to the two loop terms in the renormalization evolution of the scalar
masses is proportional to the anomalies of the $U(1)_X$  group. For the
most relevant case of  gauged $U(1)_X$ symmetries the anomalies must be
cancelled by the Green-Schwarz mechanism. We will concentrate on
theories such as the weakly coupled heterotic string with only one
anomalous $U(1)_X$ and one dilaton-axion to implement the Green-Schwarz
mechanism.

An interesting aspect of this result is that despite the large variety
of $X$ charges and choice of symmetry to explain the fermion hierarchy
the $U(1)_X$ anomaly can be fixed in a rather model independent way.
Indeed by using the previously obtained relations between the anomaly
$\cal{A}$ and the fermion masses one gets\cite{ir}
\begin{equation}
m_u m_c m_t (m_e m_{\mu} m_{\tau})^3 (m_d m_s m_b )^{-2}\approx
\left( \frac{g^{2}_{X} \cal{A} }{32\pi^2 }\right) ^{\mbox{${\cal A}/2$}}
\sin ^3\beta \cos^3\beta (174\mbox{GeV})^6 \ ,\label{mass1}
\end{equation}
where the quark and lepton masses are taken at the scale $g_X\sqrt{\cal
A}M_P/4\pi$, $\tan \beta$ is the ratio between the two Higgs vacuum
expectation values (vev's), $g_X$ is the $U(1)_X$ gauge coupling and
${\cal A}$ is the $U(1)_X$ anomaly with respect to the standard model
gauge groups.  

The evaluation of (\ref{mass1}) yields  ${\cal A}\approx 25\pm 3 .$
Remarquably enough this leads to a prediction of the Cabibbo angle
$\theta_C\approx 0.2$ in these models. 

This establishes a quite model independent estimate of the two loop
effects in inverse hierarchy models. It turns out to be much smaller
than values considered in previous discussions inspired by these
models\cite{dg,mura}.
\subsection{\sl Maximal hierarchy limit}
As already stressed in the iMSSM context, the supersymmetric flavour
problem requires very heavy sfermions in the first two generations,
hence a large  supersymmetry breaking scale, $m^2_{3/2}>>M_Z^2$.  The
fine-tuning problem becomes crucial.  In the cMSSM, where the
coefficient of the universal soft scalar masses $m_0^2$ in the
expression for $M_Z^2$ is strongly suppressed -- a Nature fine-tuning
of the top mass $m_t$ -- for moderate and large values of $\tan \beta$,
the necessary fine-tuning is mostly between $M_{1/2}^2$ and $\mu^2$.
This requires relatively large values for $\mu,$ yielding gaugino-like
LSP.  Several studies in the literature\cite{poko} have already
discussed how the predictions are modified by departing from the cMSSM
scalar degeneracy. The main point is that, in this case, the scalar
masses can also participate in cancelling the $M_{1/2}^2$ term in the
expression for $M_Z^2$, allowing for relatively low values of $\mu.$
For these small $\mu$ solutions, the spectrum wiil consist of scalars
heavier than gauginos, which are also heavier than the higgsinos.  The
advantage of this class of models is that the fine-tuning now occurs
mostly in the ratio $M_{1/2}^2/ m^2_{3/2}$, which is more obviously
related to the supersymmetry breaking mechanism, while the origin of
the $\mu$ parameter, which could now be of $O(M_Z),$ remains more
mysterious.

In Section 4, we investigate the possibility of large values of
$m^2_{3/2}$ together with small $\mu$ values. The two-loop correction,
controlled by the anomaly, contributes in some cases.  In
this regime, the iMSSM does reveal an inversed mass spectrum as compared
with the cMSSM. We discuss approximate constraints in the neighbourhood
of the 'infinitely' fine-tuned solution for large $\tan\beta,$ but this
rough approximation turns out to be quite appropriate from our numerical
analysis.

The iMSSM mass patterns are displayed in Section 5.  We summarize our
conclusions in the last section.  The case of more than one $U(1)$
flavour symmetries is sketched in the Appendix.

\section{Two Loop Renormalization Effects}

Let us first treat the two loop renormalization group equations of the
scalar masses in the following approximation,
\begin{equation}
\frac{dm_i^2}{dt}= -8\pi^2C_A(i)\left[ M_A^2 -
\frac{g_A^2}{16\pi^2}\mbox{tr}\left(\frac{C_Am^2}{d_A}\right)\right] +
2g_1^2Y_i \left[ S + \frac{g_A^2}{4\pi^2}\mbox{tr}(YC_Am^2)\right] 
\label{rge}
\end{equation}
where the indices $A=1,2,3$ correspond to the gauge group factors
$U(1), SU(2), SU(3)$ respectively, $g_A$ is the corresponding gauge
coupling, $d_A$ the algebra dimensions, $M_A$ is the gaugino mass and
$C_A(i)$ is the Casimir eigenvalue for the fermion/sfermion labelled by
$i$. Finally, 
\begin{equation}
S= \mbox{tr} (Ym^2)
\end{equation}
introduces a $Y-$dependent term in the scalar masses.

The approximation (\ref{rge}) is appropriate for the class of models
discussed herein and corresponds to neglecting the two-loop corrections
proportional to $M_A^2$ which are suppressed by a factor of
$g_A^2C_A(i)/16\pi^2$ with respect to the one loop terms. As displayed,
(\ref{rge}) is not valid for the stops. Including the top Yukawa
couplings to extend the calculation to the third family scalar masses
and consistently neglecting the terms in $M_A^2$ which are higher
order in $g_A^2$, yields the solution
\begin{eqnarray}
m_i^2&=&m_i^2(0)-\frac{a(i)}{12}\rho \left( 3\bar{m}^2 +8M_{1/2}^2+
(1-\rho)(A_{Q_3}(0) +2M_{1/2})^2-\delta \right)\nonumber \\
&+& \vert t\vert C^2_A(i)g_A^2\left (8\left (g_A^2+
\frac{1}{2}\right)M_{1/2}^2 -\delta\right) + 
\frac{Y_i}{22}\left[ S(t)-S(0)\right] \label{massren}
\end{eqnarray}
at the Fermi scale where the coefficients $a(i)$ are
$a(U_3)=4,\ a(Q_3)=2,\ a(H_2)=6$ and zero otherwise, $3\bar{m} ^2 =
(m_{U3}^2+m_{Q3}^2+m_{H2}^2),$ $A_{Q_3}$ is the soft coupling associated to
the top Yukawa coupling, and finally,
\begin{equation}
\rho=\frac{h_t}{(h_t)}_{F.P.} \approx \frac{0.72}{\sin^2\beta} \ .
\end{equation}  
In  approximating the solutions for $m_i^2$ we are anticipating and
taking advantage of the fact that the two-loop term
\begin{equation}
\delta=\frac{1}{4\pi^2}\mbox{tr}\left(\frac{C_Am^2}{d_A}\right)\vert_{t=0}
\label{delta}
\end{equation}
is almost independent of the index $A$ in the classes of models
discussed here, that we now turn to discuss.
\subsection{\sl In the cMSSM}
The most important effect of the radiative corrections on the
$SU(2)\times U(1)$ breaking appears in the Higgs parameter 
\begin{eqnarray}
m_{H_2}^2&\approx &m_{H_2}^2(0) + 0.52(M_{1/2}^2 -0.15\delta) 
-0.014 S_0 \nonumber \\
& -&\frac{0.36}{\sin^2\beta}\left(\bar{m}^2+8M_{1/2}^2
+\left( 1-\frac{0.72}{\sin^2\beta}\right)(A_0+2M_{1/2})^2
-\delta\right) \ .
\end{eqnarray}
Therefore the two loop effects due to possible heavy scalars in the
first two generations become relevant for $\delta \sim O({\rm
few}M_{1/2}^2).$ For instance assuming a degeneracy amongst the heavy
scalars of mass $m$ in the first two generations this corresponds to $m
\sim 5 M_{1/2}.$

It is well known that in the cMSSM with boundary conditions
$m_i^2(t=0)=m_0^2$ the coefficient of $m_0^2$ in $M_Z^2$ is small for
large values of $\tan \beta$. Indeed the dependance of $m_{H_2}^2$ on
$m_0^2$ in the $\tan\beta >> 1$ limit is
\begin{equation}
m_{H_2}^2\approx -0.1 m_0^2 +0.3 \delta -2.76 M^2_{1/2} \ .
\end{equation}
Taking the traces over all the MSSM scalars one gets
$\mbox{tr}(C_3/8)=6,\ \mbox{tr}(C_2/3)=7,\ \mbox{tr}C_1=6.6,$ 
so that
\begin{equation}
\delta\approx \frac{m_0^2}{2\pi} \label{deluni} \ .
\end{equation}
The possibility of obtaining  a relatively large value of $m_0^2$
without worsening the fine-tuning in $M_Z^2$ remains at the two-loop
level.

\section{Anomalous U(1) Models} 

As already emphasized in the introduction the natural realization of
the inverse hierarchy for squarks and sleptons occurs in models where
the fermion mass hierarchy is explained by a Frogatt-Nielsen mechanism
with an anomalous $U(1)_X$ local flavour symmetry. Recently it has been
advocated that in type IIB orientifold models of string theory one
could expect different anomalous  $U(1)_X$ with their anomalies being
cancelled by a corresponding number of moduli fields \cite{aqi}.  We
shall ignore this possibility and remain within the more traditional
heterotic-like picture with only one anomalous $U(1)_X$ \cite{dws}. Of
course one could postulate the existence of other anomaly-free $U(1)$
flavour gauge symmetries in order to explain the fermion mass
hierarchy. As discussed in the Appendix the results are essentially
similar in the multi-$U(1)$ models.

We refer to the comprehensive literature on this subject for the
details and quote the main results only. The anomaly cancellation is
provided by the Green-Schwarz mechanism. A necessary condition for the
compensation of the anomalies with respect to the standard model gauge
symmetries as well as the gravitational anomaly is the equality
\begin{equation}
{\cal A}_1= {\cal A}_2={\cal A}_3=\frac{\mbox{tr}(X)}{24}\ ,\label{Acond}
\end{equation}
where ${\cal A}_A= 2\mbox{tr}(XC_A)/d_A .$ The Kac-Moody levels $k_A$
have all been taken to be one to simplify the discussion. The
coefficiient of the Fayet-Iliopoulos term required by the $U(1)_X$
gauge invariance -- which is nothing but the contribution of the
dilaton to $D_X$ -- is given by
\begin{equation}
\xi^2= \frac{k_X g_X^2 {\cal A}}{32\pi ^2} \ .\label{xi} 
\end{equation}
Introducing a Froggat-Nielsen field $\phi$ which is standard
model gauge singlet with a charge $X=-1$ (by a suitable
normalization of the $U(1)_X$ charges), the $U(1)_X$ gauge
symmetry is broken at the minimum of the scalar potential where
\begin{eqnarray}
\vert \phi \vert ^2&=&\xi^2 M_P^2 \ , \nonumber \\
\left<F_{\phi}\right> &\sim &m_{3/2}\xi M_P \ ,\nonumber \\
\left<D_X\right> &=&m_{\phi}^2 \ ,\label{breaking}
\end{eqnarray} 
where $ m_{\phi}^2$ is the soft mass given by the supersymmetry
breaking mechanism.  The $\left< F_{\phi}\right>$ and $\left<
D_{X}\right>$ vev's are  the induced supersymmetry breaking terms. The
former contributes to the scalar masses only as $\xi^2m_{3/2}^2$ and is
neglected here\footnote{Though, as discussed in \cite{dps}, this
contribution is relevant in the discussion of FCNC effects.}. The latter
gives an important D-term contribution to the scalar masses so that at
the scale $\xi M_P$,
\begin{equation}
m_i^2=\hat m_i^2 +X_i \left< D_X\right> \ ,\label{XD}
\end{equation}
where $\hat m_i^2$ is the contribution from the supersymmetry breaking
independent of the $D_X$ breaking and the charges $X_i$ are model
dependent.
\subsection{\sl Two-loop Correction and Anomaly}
In the inverse hierarchy models the two loop contributions of the
heavy scalars to the renormalization group equations turns out to
be quite independent of the choice of charges. Indeed, one gets
from the definition (\ref{delta}) and the masses (\ref{XD}),
\begin{equation}
4\pi^2\delta =\sum_i \frac{C_A(i)}{d_A}(\hat m_i^2 +X_i<D_X>) \ ,
\end{equation}
which yields
\begin{equation}
\delta=\hat \delta +\frac{1}{8\pi^2}{\cal A}\left< D_X\right>\ . 
\label{delA}
\end{equation}
In obtaining this result we have used the equality of the anomalies
(\ref{Acond}). Hence, the main two-loop contribution, coming from the
$D-$terms in the scalar masses, is proportional to the anomaly
${\cal A}$. Of course, the two-loop corrections coming from $D-$terms
corresponding to non-anomalous $U(1)'$s cancel.   

Interestingly enough, the anomaly ${\cal A}$ can be calculated from its
relation to the fermion masses. Indeed, even if the $X-$charges that
control the fermion masses are model dependent, it is possible to
display a combination of masses that only depends on the charges
through ${\cal A,}$ as we now turn to discuss.
\subsection{\sl Calculation of the Anomaly}
In the anomalous $U(1)_X$ approach to the fermion hierarchy, with the
abelian flavour symmetry breaking given by the small parameter $\xi$
as discussed above, the quark and lepton Yukawa couplings to the
Higgses are given by
\begin{equation}
Y_{f_i} \sim  \xi ^{f_{Li} + f_{Ri} + h} \ ,\label{Yuky}  
\end{equation}
where the fermion name ({\sl e.g.}, $q_{i}$) also denotes its
$X-$charge (resp., $X(u_{Li})=X(d_{Li})),$ and $h$ is the $X-$charge of
the appropriate Higgs field. In particular, one obtains, from the
values of the third generation Yukawa couplings, the relations:
\begin{eqnarray}
q_3 + u_3 + h_2 &\approx & 0 \ ,\nonumber \\ 
q_3 + d_3 + h_1 &\approx & l_3 + e_3 + h_1 \approx \frac{4 - 
\ln{\tan \beta}}{\ln{\xi}} \ ,\label{charges} 
\end{eqnarray}
and $\xi$ is roughly the Cabibbo angle, $\theta_{C}\approx \xi.$
The charges of the other family fermions are more or less fixed by
their Yukawa couplings and the Kobayashi-Maskawa mixings.

Because of the condition (\ref{Acond}) on the anomalies, ${\cal A}$
is related to the fermion masses as follows,
\begin{eqnarray}
m_um_cm_t\left( m_em_{\mu}m_{\tau}\right) ^3(m_dm_sm_b)^{-2}\approx
\xi ^{\mbox{${\cal A}$}}\sin ^3\beta \cos^3\beta (174\mbox{GeV})^6 \ ,
\label{calA}
\end{eqnarray}
where $\xi^2=\frac{g_X^2{\cal A}}{32\pi^2}$, while for the Higgs
charges one obtains,
\begin{equation}
(h_1 + h_2) \ln{\xi} \approx 
\ln \left( \frac{m_d m_s m_b}{m_e m_{\mu}m_{\tau}} \right) \ .
\label{higgs}
\end{equation}

The quark and lepton masses are defined at the scale $\xi M_P .$ 
Putting the experimental masses in (\ref{calA}) yields 
\begin{eqnarray}
{\cal A}\ln(\frac{32\pi^2}{g_X^2{\cal A}})
\approx (90 \pm 3\ln{\sin{2\beta}} ) \ ,\label{Aeq}
\end{eqnarray}
giving ${\cal A}\approx 25\pm 3$ and $\xi \approx 0.2,$ with the GUT
value $g_X^2 = .5.$ This is in reasonable agreement with the relation
$\theta_C\approx \xi$. From (\ref{higgs}) one gets $(h_1 + h_2) \approx
0 ,$ a result that we shall use later. In the Appendix we discuss the
model dependence of these relations.

Therefore, as a typical result, the relevant two-loop contribution 
to the low-energy scalar masses (\ref{massren}) is given by  a
relatively low value, 
\begin{equation}
\delta \approx \frac {m_{\phi}^2}{\pi}+ \hat{\delta} \label{delgen}
\end{equation}
where we have used (\ref{breaking}). 
\subsection{\sl i+cMSSM }
Let us first evaluate the impact of this two loop correction in a
simple model (i+cMSSM) with universality assumed for the primordial
supersymmetry breaking, namely, a contribution $m_0^2$ to all scalar
soft masses, and with an anomalous $U(1)_X$ flavour symmetry as
discussed above. In this case, $\left< D_X\right> = m_0^2,$ and the
scalar masses at the flavour symmetry breaking scale are
\begin{equation}
m^2_i=m_0^2 \left( 1 + X_i \right).
\end{equation}
From (\ref{charges}), the parameter $\bar{m}^2$ of the one-loop
correction in (\ref{massren}) is equal to $m_0^2,$ and, from
(\ref{delgen}) and (\ref{deluni}), the two-loop contribution depends on
\begin{equation}
\delta\approx {m_0^2}\left( \frac{1}{2\pi} + \frac{1}{\pi} \right)
\approx 0.5 m_0^2 \label{cdel}
\end{equation}
The corresponding contribution to $m_{H_2}^2$ is
\begin{equation}
\Delta m_{H_2}^2 \approx \frac{0.2 m_0^2}{\sin^2\beta}
\end{equation}
which is about three times larger than the corresponding parameter in
the cMSSM. As we shall discuss in the next section, contrarily to what
happens in the cMSSM, this alone could be enough to allow for small
$\mu$ models, in the large $\tan{\beta}$ limit, even if $X(H_2 ) = 0.$

\subsection{\sl iMSSM}
We now turn to discuss a more elaborate model\cite{dps} exhibiting the
inverse hierarchy for the scalar masses, where some flavour dependence
is incorporated into the K\"{a}hler potential besides the anomalous
$U(1)_X$ gauged flavour symmetry and the Green-Schwarz mechanism.  The
key assumption, which allows  the model to be  predictive, is that the
{\sl fermion mass hierarchies} are fixed solely by the abelian flavour
symmetries, not by the moduli dependence.  Amazingly, this assumption
turns out to imply that the (mass)$^2$ differences between sfermions
with the same SM quantum numbers are proportional to the gravitino
(mass)$^2,$ $m_{3/2}^2.$ Even without specifying the primordial
supersymmetry breaking along the dilaton and moduli directions in the
iMSSM, the sfermion masses can be parameterized in a simple and
suggestive way, which we now turn to summarize.

Let us denote by $\Phi_i$ the matter superfields and their scalars
where $\Phi=Q,U,D,L,E$ refers to the standard model fields and
$i=1,2,3$ refers to the family index. We denote by $\phi_i$ the $X$
charges. At the scale $\xi M_P$ the soft terms satisfy the relations
\begin{eqnarray}
m_{\Phi_i}^2-m_{\Phi_j}^2&=&(\phi_i-\phi_j)m_{3/2}^2, \nonumber \\  
m_{U_3}^2+m_{Q_3}^2+m_{H_2}^2&=&M_{1/2}^2,\nonumber \\
m_{D_3}^2+m_{Q_3}^2+m_{H_1}^2&=&M_{1/2}^2+(d_3+q_3+h_1)m_{3/2}^2,\nonumber\\
m_{E_3}^2+m_{L_3}^2+m_{H_1}^2&=&M_{1/2}^2+(e_3+l_3+h_1)m_{3/2}^2,\nonumber\\
m_{H_2}^2 + m_{H_1}^2&=&(2+h_2+h_1)m_{3/2}^2,\nonumber \\ 
A_{U_i}&=&(u_i+q_i+h_2 )m_{3/2}-M_{1/2}\nonumber \\
A_{D_i}&=&(d_i+q_i+h_1)m_{3/2}-M_{1/2}\nonumber \\
A_{L_i}&=&(e_i+l_i+h_1)m_{3/2}-M_{1/2}\nonumber \\ 
B&=&\left( 2+(h_2+h_1)\theta(h_2+h_1)\right) m_{3/2} \label{dpsmodel}
\end{eqnarray}
Actually, the terms proportional to the $U(1)_X$ charges come out as a
particular combination of the $\left< D_X\right>$ induced breaking and
the supersymmetry breaking in the moduli sector, which give rise to
this general form for the mass splitting between families. We have only
considered the soft terms which are diagonal in the family indices,
although the pattern of the off-diagonal terms give constraints on the
$U(1)_X$ charges from the FCNC bound. Other relations for the soft
terms will be spelt out later.

The fermion hierarchy requires a relatively strong family ordering of
the charges $\phi_1>\phi_2 >\phi_3$.  We concentrate on this situation
and even more on the case where $M_{1/2}<m_{3/2}$. However we do not
have to impose any particular choice for the charges $\phi_i$, in many
of the physical issues discussed below since the two loop corrections
that are relevant to the inverse hierarchy scenario are controlled by
the anomaly. This is fixed by the fermion masses to
\begin{equation}
\delta =\frac{{\cal A}m_{3/2}^2}{8\pi^2}\approx \frac{m_{3/2}^2}{\pi}\ .
\label{deli}
\end{equation}
With the relations in (\ref{dpsmodel}) one has for instance
\begin{equation}
m_{H2}^2=m_{H2}^2(t=0) -\frac{0.36}{\sin^2\beta}\left(\left( 10-
\frac{0.72}{\sin^2\beta}\right) M_{1/2}^2 -\delta\right) 
+0.52( M_{1/2}-0.15\delta)
\end{equation}
Therefore, the two-loop corrections are basically negligible in
the iMSSM. Nevertheless, the $\delta$ term is of some importance in
the discussion of the next section.  
\section{Higgsino-like LSP and Electro-Weak Symmetry Breaking}

It is well-known that by departing from the universality assumptions of
the cMSSM many of its striking predictions are dramatically affected.
In particular the nature of the LSP can change. This is mainly a matter
of competition between  the $\mu^2$ and $M_{1/2}^2$ parameters that
appear with different signs in the supersymmetric expression for
$M^2_Z$. The cMSSM coefficient of the universal soft scalar masses
$m_0^2$ is strongly suppressed  for rather large values of $\tan
\beta$. The necessary fine-tuning between $M_{1/2}^2$ and $\mu^2$ then
favours a lighter gaugino than the Higgsino.  In this section we show
that in the iMSSM discussed in the previous sections, the Higgsino
turns out to be a natural option for the LSP. Let us  sketch 
the situation within an analytic approximation to the supersymmetric
$SU(2)\times U(1)$ breaking. In terms of $t=\tan \beta, $ the minimum
equations read
\begin{eqnarray}
m_{H_1}^2-m_{H_2}^2-B\mu t(1-\frac{1}{t^2})=M_Z^2
\frac{1-t^2}{1+t^2}\nonumber \\
m_{H_1}^2+m_{H2}^2 + 2\mu^2-B\mu t(1+\frac{1}{t^2}) =0
\label{minimum} 
\end{eqnarray}
at the classical level. The radiative corrections are important
but the main contributions can be included by redefining
\begin{eqnarray}
\hat{m_1}^2 &=& m_{H_1}^2 -\frac{M_Z^2}{2}\frac{1-t^2}{1+t^2}
+3\frac{h_t^2}{16\pi^2}\mu^2 \ln \left(\frac{m_{\tilde t_1}
{m_{\tilde t_2}}}{m_t^2}\right) \nonumber \\
\hat{m_2}^2 &=& m_{H_2}^2 +\frac{M_Z^2}{2}\frac{1-t^2}{1+t^2}
+3\frac{h_t^2}{16\pi^2}(m_{\tilde t_1}^2+m_{\tilde
t_2}^2)\ln \left(\frac{m_{\tilde t_1}{m_{\tilde
t_2}}}{m_t^2}\right) \ . \label{newvariable}
\end{eqnarray}
From the minimum equations  we deduce that
\begin{equation}
\mu=\frac{Bt}{2}(1\pm \sqrt{1-\frac{4\hat{m}_{1}^2}{B^2 t^2}}) \ ,
\end{equation}
so that the small $\mu$ solution leading to a Higgsino-like LSP
corresponds to the minus sign in the previous equation. For large
enough values of $t$ (to be discussed later) such that $0<
4\hat{m}_{1}^2 << B^2t^2$ one gets the following relations
\begin{eqnarray}
\mu&\approx &\frac{\hat{m}_{1}^2}{Bt}(1+\frac{\hat{m}_{1}^2}{B^2t^2})
\nonumber \\
\hat{m}_{2}^2&\approx &\frac{\hat{m}_{1}^2}{t^2}
(1-\frac{\hat{m}_{1}^2}{B^2}) \label{master}
\end{eqnarray}
Namely the small $\mu$ regime corresponds to $\mu^2 \propto t^{-2}$ and
$m_{H_2}^2\propto t^{-2}$. This suggests to expand the solutions in
powers of $t^{-2}$. This is quite unphysical but mathematically sound.
Let us start with an approximation to the low energy masses
parameterized as follows
\begin{eqnarray}
m_{H_1}^2&=&m_0^2(1-\sigma_H)+\frac{1}{2}(M_{1/2}^2 -0.15\delta)
\nonumber \\
m_{H_2}^2&=&m_{H1}^2 +2\sigma_H m_0^2 
- 0.36\left( 1+\frac{1}{t^2}\right) \left(\bar m^2
+8M_{1/2}^2+\Delta-\delta\right) \label{m_H}
\end{eqnarray}
where
\begin{eqnarray}
\Delta = \left( 0.28-\frac{0.72}{t^2} \right)\left(A_{U_3}
+2M_{1/2}\right)^2
\end{eqnarray}
is model dependent. In (\ref{m_H}), $m_0$ is the universal scalar
mass in the cMSSM and $m_0 = m_{3/2}$ in the iMSSM, as discussed before,
and we have assumed $m_{H_1}^2+m_{H_2}^2=2m_0^2$
as consistent with $h_1+h_2=0$.

As a first approximation we determine $M_{1/2}^2 / m_0^2$ by taking the
large fine-tuning limit, $M_Z^2<<m_0^2.$  We also neglect the radiative
corrections and we keep only the relevant powers of $\tan \beta.$ Then,
one can solve (\ref{master}) for $M_{1/2}$ in each of the models
discussed in the previous sections.  For the sake of illustration, we
take the values predicted by the iMSSM, $A_{U_3} = -M_{1/2}$ and $B =
2m_0,$ but the latter only enters into the term $\propto t^{-2} .$
\vskip .2 cm
\leftline{{\bf a) cMSSM}}
\vskip .2 cm
\noindent
In this model, $\sigma_H=0,\ \bar m^2=m_0^2,\ \delta\approx
m_0^2/(2\pi) .$ One gets
\begin{equation}
\frac{M_{1/2}^2}{m_0^2}\approx
-0.02 - \frac{0.7}{t^2}
\end{equation}
which excludes the small $\mu$ solution in the limit $m_0^2
>>M_Z^2$ as well-known.
\vskip .2 cm
\leftline{{\bf b) i+cMSSM}}
\vskip .2 cm
\noindent In this case, the universality is only broken by the $U(1)_X$
$D-$terms, so that $\bar m^2=m_0^2 $ from (\ref{charges}),
$\delta\approx 3m_0^2/(2\pi)$ from (\ref{cdel}), $\sigma _H = h_2,$
yielding, 
\begin{equation}
\frac{M_{1/2}}{m_0^2}\approx  0.04 + 0.4 \sigma_H 
-\frac{0.7+0.3\sigma_H}{t^2}
\end{equation}
The existence of this solution, especially for $h_2=0$, is due to the
larger two-loop contributions. The gauginos are much lighter than the
sfermions for these small $\mu$ solutions. In this example, a
cancellation must occur between the one-loop term in $M_{1/2}^2$ and
the two-loop term in $m_0^2$ for large values of the soft masses.
\vskip .2 cm
\leftline{{\bf c) iMSSM}}
\vskip .2 cm
It is characterized by $\bar m^2=M_{1/2}^2,$ from (\ref{dpsmodel}),
and $\delta \approx m_{3/2}^2/\pi ,$ from (\ref{deli}). This leads to
\begin{equation}
\frac{M_{1/2}^2}{m_{3/2}^2}\approx 0.36(1+\sigma_H) + .05
-\frac{0.7+0.3\sigma_H}{t^2} \label{rapp}
\end{equation}
Roughly, the parameter $\sigma _H $ can take values in the range [-1,0].
{\sl E.g.,} if we take the value $\sigma _H = -0.25$  and $t=2$, we get
$0.17$ for the ratio (\ref{rapp}), which is close to the values
obtained in a scanning of the parameter space. Notice that the two-loop
(anomaly) term contributes by about one-third to this result. The ratio
in (\ref{rapp}) means a real fine-tuning, and in our numerical analysis
(after reintroduction of the radiative corrections and $M_Z$ in the
expressions) the deviations from this `infinite fine-tuning' limit are
rather small. In order to allow for a big hierarchy in the sfermion
masses, one has to take rather small values of $M_{1/2}/m_{3/2} ,$ by
increasing $|\sigma_H | .$ This ratio is related to the Goldstino angle
$\sin\theta_G= M_{1/2}/\sqrt{3}m_{3/2}$. In a sense this is a better
variable to be tuned than $\mu/m_{3/2}$ since it is simply related to
the nature of the supersymmetry breaking. Still it has to be fine-tuned
to match a quantity which, in the iMSSM, depends on the parameter
$\sigma_H$, related to the properties of the Higgs fields under the
$U(1)_X$ and modular symmetries.

Notice that the condition for a Higgsino-like LSP, $\mu^2<M_{1/2}^2/6
,$ is fulfilled with $\tan \beta > 3,$ for $\sigma _H > -.75 .$
Otherwise, the LSP can be gaugino-like, while the lightest chargino
remains Higgsino-like.  Therefore, the small $\mu$ solution of the
iMSSM is generically characterized by {\sl (i)} large values of
$m_{3/2},$ {\sl i.e.,} very heavy sfermions of the two first
generations; {\sl (ii)} smaller values of $M_{1/2},$ {\sl i.e.,}
moderate gaugino masses; and {\sl (iii)} $\mu$ as low as $O(M_Z),$ {\sl
i.e.,} a Higgsino as the LSP.

Radiative corrections have been neglected in this discussion, but the
main effect of their inclusion is to increase the value of $\tan \beta$
for a given set of parameters.

\section{The iMass Spectrum}

In this section we shall discuss the typical mass spectrum that one can
derive from inverse hierarchy models. We will also comment briefly on
the predictions for the FCNC effects, a main issue in these models
because of the large mass splitting between the families.

The mass spectrum of the iMSSM version\cite{dps}  depends on three
parameters $\sigma_H$, $\sigma_L$ and $\sigma_Q$ which measure the
departure from scalar mass universality between the two Higgs
doublets,  the leptons and the quarks of the third family,
respectively.  These parameters define the solutions of
(\ref{dpsmodel}) and so include the dependence on the corresponding
$X-$charges and, in this model, on their flavour dependent K\"{a}hler
geometry. They are family independent. Then, the family dependent mass
terms, accordingly to (\ref{dpsmodel}), depend only on the $X-$charge
differences, {\it e.g.,} $q_{1}-q_{3}.$  Such charges can be chosen to
get a good agreement with fermionic mass patterns and mixing angles.
The choice of charges plays also a role in the $S$ term in the masses
(at the Fermi scale). The correction to the masses due to this term is
\begin{equation}
\delta m_i^2 =\frac{Y_i}{22}(S-S_0)  \ ,\label{Sterm}
\end{equation}
where $S_0$ comprises a term like $\hbox{tr} (XY) m_{3/2}^2 .$ From the
renormalization group equations one finds that the evolution of $S$ is
given in first approximation by
\begin{equation}
S-S_0=(\frac{g_1^2}{g_0^2}-1)S_0 \approx -0.6 S_0 \ .
\end{equation}
The effect of this contribution has been often overestimated in the 
literature. In any instance, the term (\ref{Sterm}) can be consistently 
included in the definition of $\sigma_H$, $\sigma_L$ and $\sigma_Q ,$
without loss of  generality. This is understood in what follows.

The masses of the third family sleptons are then given by $(\kappa=
M_Z^2\cos 2\beta )$
\begin{eqnarray}
m^2_{\tilde\tau_L}&=&(1+\sigma_l)m_{3/2}^2 +0.5(M_{1/2}^2-0.15\delta)+
0.4\kappa \nonumber \\
m^2_{\tilde\nu_L}&=&(1+\sigma_l)m_{3/2}^2 +0.5(M_{1/2}^2-0.15\delta)+
-0.5\kappa \nonumber \\
m^2_{\tilde\tau_R}&=& 1.16 M_{1/2}^2 -(\sigma_l-\sigma_H)m_{3/2}^2 
-0.03\delta +0.23\kappa \label{sleptons}
\end{eqnarray}
and the third family squark masses are given by
\begin{eqnarray}
m^2_{\tilde b_L}&=& (1+\sigma_q)m_{3/2}^2 +6.9M_{1/2}^2 -0.5\delta
-\frac{\rho}{6}((10-\rho)M_{1/2}^2-\delta)+0.42 \kappa\nonumber\\
m^2_{\tilde b_R}&=&-(\sigma_q-\sigma_H)m_{3/2}^2 +7.4M_{1/2}^2
-0.43\delta +0.75 \kappa\nonumber\\
m^2_{\tilde t_L}&=& (1+\sigma_q)m_{3/2}^2 +6.9M_{1/2}^2 -0.5\delta
-\frac{\rho}{6}((10-\rho)M_{1/2}^2-\delta)-0.35 \kappa\label{squarks}\\
m^2_{\tilde t_R}&=&-(2+\sigma_q+\sigma_H)m_{3/2}^2 +7.4M_{1/2}^2
-0.44\delta -\frac{\rho}{3}((10-\rho)M_{1/2}^2-\delta)-0.15 \kappa
\nonumber
\end{eqnarray}

Since we are  interested in the case $ M_{1/2} << m_{3/2}, $ the
allowed range for the parameters $\sigma_H$, $\sigma_L$ and $\sigma_Q
,$ is strongly constrained, and $\sigma_H$ has also  to be
consistent with the electroweak break conditions (\ref{master}).

Let us now present some generic features of the spectrum. The
differences in the charges between the first two generations and the
third one are model dependent to some extent. However, one can minimize
these uncertainties by considering the combinations that are more
directly related to the fermion masses, as given by (\ref{Yuky}).  For
the first family sfermions, as compared to the third family ones, one
finds,
\begin{eqnarray}
m^2_{\tilde{e}_L}+m^2_{\tilde{e}_R} -m^2_{\tilde\tau_L}-m^2_{\tilde\tau_R}
&\approx& \frac{\ln (m_e / m_{\tau} )}{\ln \xi}m_{3/2}^2 \, \nonumber \\
m^2_{\tilde{d}_L}+m^2_{\tilde{d}_R} -m^2_{\tilde{b}_L}-m^2_{\tilde{b}_R}
&\approx& \frac{\ln (m_d / m_{b} )}{\ln \xi}m_{3/2}^2 \, \nonumber \\
m^2_{\tilde{u}_L}+m^2_{\tilde{u}_R} -m^2_{\tilde{t}_L}-m^2_{\tilde{t}_R}
&\approx& \frac{\ln (m_u / m_{t} )}{\ln \xi}m_{3/2}^2 \ . \label{sfirst}
\end{eqnarray}
These are high energy relations that remain valid at low energies as
far as the masses of the third generation are taken from
(\ref{sleptons}) and (\ref{squarks}) without the terms proportional to
$\rho .$ Analogous expressions hold for the second generation of
sfermions.  Of course, parity conjugated sfermions are usually splitted
by a large amount with respect to the above averages. If we take, as an
example, $\sigma_H \approx \sigma_L\approx \sigma_Q \approx -1 ,$ which
leads to relatively light charginos (without further motivation for
this particular choice), all the third family sfermions are as light as
the gauginos, the first and second families are much heavier. The
two-loop contributions are important in this case where the fine-tuning
between $ M_{1/2}$ and $m_{3/2}$ is large.

\begin{figure}
\epsfxsize=12.cm
$$\epsfbox{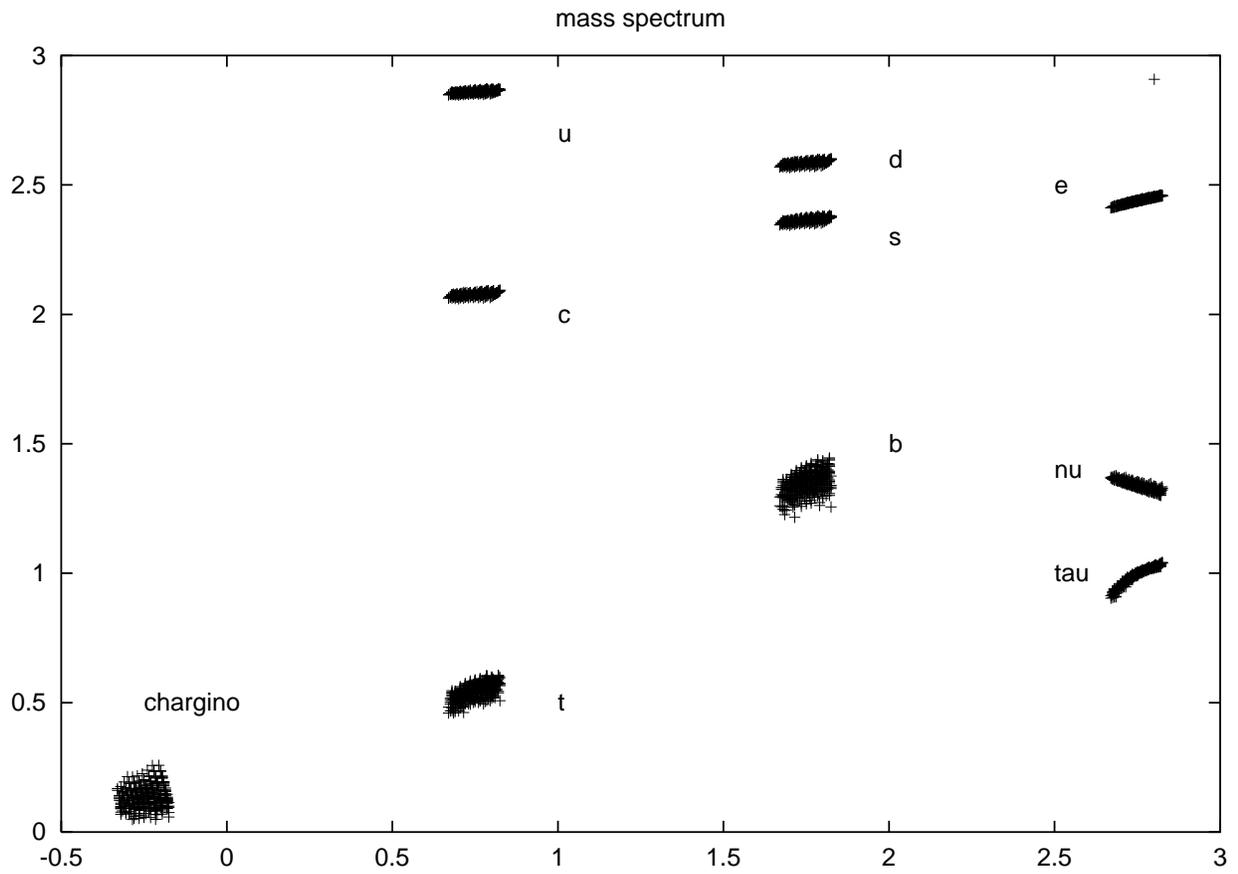}$$
\caption{The mass spectrum in units of $m_{3/2}$. We have displayed the 
lightest chargino mass and the left-right average masses for the $U$ squarks, 
the $D$ squarks and the sleptons as a function of $\sigma_H$
 appropriately shifted. }
\end{figure}

In the numerical analysis, the radiative corrections are included, and
the parameter space is scanned around the maximal fine-tuning values.
As expected we find Higgsino-like LSP's degenerate with  the lightest
chargino. The chargino masses can be as low as $100$ GeV.  We have
explicitly cut the spectrum by (arbitrarily) imposing that the MSSM
Higgs mass is greater than $100$ GeV. We do find Higgses within the
$100-109$ GeV slot corresponding to a value of $\tan \beta$ which does
not exceed $18$. Among the squarks the left sbottoms are the lightest.
We present in fig. 1 the mass spectrum as a function of $ \sigma_H$. We
have rescaled the masses and display them in units of $m_{3/2}$.  As
expected the hierarchy between families is not destroyed by the
evolution down to the Fermi scale. The values of $m_{3/2}$ chosen in
the figure are below $2$  TeV.  Higher values of $m_{3/2}$ would not
modify the picture, only the fine-tuning would be more severe.

Let us come back to one of the issues which prompted our study: the
FCNC effects and the decoupling of the first two families. The mass
insertions that are usually used to evaluate the FCNC
contributions\cite{mass} are roughly given in terms of the $X$ charges
of the particles by
\begin{equation}
\delta_{ij}\sim 2\frac{\vert X_i -X_j \vert }{X_i + X_j}
\xi^{\vert X_{i}-X_{j}\vert }
\end{equation}
The strong constraints on the mass insertions  with $i=1$ and $j=2$,
suggests\cite{dps} a choice of some degeneracy and some alignment in
the diagonal soft masses by choosing $d_1=d_2$ and $e_1=e_2$.
However, this is not enough and we still need large values of
the supersymmetry breaking parameter, $m_{3/2} \sim 2{\rm TeV}$ for a
sufficient FCNC decoupling.  Indeed, as noticed before, the flavour
dependence of the soft terms coming from the $\left< F\right>$
supersymmetry breaking have also to be taken into account. They are
reduced by at least a factor $\xi ^2,$ as follows from
(\ref{breaking}), and more model dependent, but still dangerous enough
for the $K-\bar{K}$ system. Fortunately we do get such high values of
$m_{3/2}$ in our numerical scanning without further effort. Yet, the
contribution to $\epsilon _K$ comes out close to the phenomenological
bounds in this model, in spite of the combined use of all three
anti-FCNC mechanisms, degeneracy and alignment from the equality of
some charges, together with decoupling through a relatively large
supersymmetry breaking scale in the scalar sector. This is the price to
pay for the inverse sfermion mass hierarchy. 

It is worth noticing that the small $ M_{1/2}/m_{3/2}$ and $ A/m_{3/2}$
ratios that characterize these models are what is needed\cite{abe} to avoid 
charge and colour breaking vacua without need for further cosmological
assumptions. In the large $\mu$ version of the model, one can reduce the
hierarchy and increase the degeneracy by increasing the $ M_{1/2}/m_{3/2}$
ratio. Besides the fact that it would bring back the issue of a fine-tuning
of the $\mu / M_{1/2}$ ratio, this would be strongly constrained by the
wrong vacuum problems.

\section{Summary and Concluding Remarks}
We have studied the decoupling of the first two squark families in
order to lower the FCNC effects. This has been done using the gauged
$U(1)$ flavour symmetries which had already been utilised to explain the
fermion masses  and the mixing angles. Within this framework we have
focused on more model-independent results. In particular as soon as one
tries to induce large mass hierarchies one faces the fine tuning
problem in the electroweak sector. Indeed the Fermi scale has to be
maintained although the supersymmetry breaking scale is pushed up
beyond the TeV limit.  This forces to study carefully the diverse
compensations in the $M_Z^2$ equation. As a result one has to resort to
a two-loop analysis of the Higgs sector. Fortunately we have shown that
the two-loop effects are solely governed by the Green-Schwarz anomaly
which is determined from the fermion masses. This allows a thorough
study of the minimum equations, and the possibility of a scenario
where  the fine-tuning appears in the $(M_{1/2},m_{3/2})$ sector with a
small value for $\mu$.  This differs from the usual cMSSM where $\mu$
is large. This leads to a Higgsino-like LSP and a characteristic mass
spectrum. In particular we find light charginos. On the contrary, the
sfermions of the first and second families should be of order a few TeV,
a nice experimental signature, indeed.

Of course, the inverse hierarchy models based on abelian flavour
symmetries are especially affected by the FCNC problems. The
supersymmetry breaking scale required to get an efficient decoupling
would be very high. Therefore, it is not clear whether they are a good
choice to escape the flavour changing effects in spite of the natural
prediction of heavy sfermions in the first two families. On the other hand they also
possess some other nice features: a natural small scale from the
Fayet-Iliopoulos term, the presence of anomalous abelian symmetries in
superstring solutions, the simplicity of the fitting to the puzzling
fermion hierarchy and, last but not least, the relation between the
fermion and sfermion spectrum. A compromise could be obtained with
additional flavour symmetries, for instance non-abelian ones, to reduce
the splitting between the sfermion in the first two families. More
speculatively, one could hope that the more recent developments in
string theory -- see for instance \cite{aqi, alda} and references
therein -- would provide new insights into the old quarrel of
supersymmetry with flavour. 
   
\newpage
\section{Appendix}
In the case of more than one abelian flavour charges, $X_i,$ $(i=1,....,
n),$ we introduce an equal number of scalars, $\Phi ^i,$ so that all
the $U(1)$'s are broken. For the consistency of the model, we make the
following assumptions:

\noindent A) Only one $U(1)$ symmetry is anomalous, which we call
$X_1,$ and only the corresponding $D-$term has a Fayet-Iliopoulos term
with coefficient $\xi .$ This is related to the anomaly $\cal{A}$
through the Green-Schwarz mechanism by (\ref{xi}). It is mandatory that
the other abelian charges fulfil the analogous of (\ref{Acond}) and
that they do not introduce any other anomaly.

\noindent B) Let us denote by $-\phi _{ij}$ the charge $X_i$ of the
scalar $\Phi ^j$. They are chosen so that there is no term in the
superpotential with the $\Phi '$s alone. These $U(1)$ charges are
normalized so that all the corresponding coupling constants are equal.

The relevant soft-terms are the masses $m_i$ of the scalars $\Phi ^i.$
Let $\phi ^{-1}_{\  ij}$ be the inverse of the charge matrix $\phi$
defined above, which has an inverse because of our assumption B). The
equivalent of (\ref{breaking}) is now,
\begin{eqnarray}
\vert \phi_i\vert ^2&=& \phi ^{-1}_{\  i1}\xi^2 M_P^2, \nonumber \\
\left<D_{X_i}\right> &=&m_{j}^2 \phi ^{-1}_{\  ji}\label{breakingnew}
\end{eqnarray} 
Then, (\ref{calA}) is modified by the replacement,
\begin{eqnarray}
\xi ^{\mbox{${\cal A}$}}\ \ \longrightarrow \ \  \prod \left( \phi ^{-1}_{\  i1}\xi^2
\right)^{\mbox{${\cal{A}}_i /2$}} , \label{replace}
\end{eqnarray} 
where ${\cal A}_i = \phi ^{-1}_{\  i1}{\cal A}.$ Therefore, in the
pluri-$U(1)$ case, the resulting value for the anomaly can be slightly
different  from the value in section $3$.

Finally, the contributions to the sfermion masses from the
$\left<D_{X_i}\right>$ terms are $ m_{j}^2 \phi ^{-1}_{\  ji}X_i (a)$,
where $X_i (a)$ is the corresponding fermion charge. The contribution
to the two-loop scalar masses becomes,
\begin{eqnarray}
\delta=\hat \delta +\frac{1}{8\pi^2}m_{j}^2 \phi ^{-1}_{\  j1}{\cal A}
\label{deltamulti}
\end{eqnarray}
where $\delta$ is defined in (\ref{delta}). This allows for some
variation with respect to the values discussed in section $3$, but the
two-loop contributions generically remain as small.


\begin{thebibliography}{99}

\bibitem{nls}
M. Leurer, Y. Nir, N. Seiberg,  {\it Nucl. Phys.} {\bf B398} (1993) 319;
{\bf B420} (1994) 468;
Y. Nir and N. Seiberg, {\it Phys.\ Lett. }{\bf B309 }(1993) 337.

\bibitem{dg}
S. Dimopoulos; G.F. Giudice, {\it Phys. Lett.} {\bf 357B} (1995) 573;
A. G. Cohen, D. B. Kaplan, A. E. Nelson, {\it Phys. Lett.}{\bf 388B}
588 (1996);
A. Pomarol, D. Tommasini, {\it Nucl.\ Phys. }{\bf B466 }(1987) 3.

\bibitem{mura} 
N. Arkani-Hamed, H. Murayama, {\it Phys. Rev.}{\bf D56 }(1997) 6733;
K. Agache, M. Graesser, {\it Phys. Rev.}{\bf D59 }(1999) 015007.

\bibitem{FN}  
C. D. Froggatt and H. B. Nielsen, {\it Nucl.\ Phys. }%
{\bf B147 }(1979) 277;
J. Bijnens and C. Wetterich, {\it Nucl.\ Phys. }{\bf B283 }(1987) 237.
P. Ramond, R.G. Roberts and G. G. Ross, {\it Nucl.\
Phys.\ }{\bf B406 }(1993) 19.

\bibitem{ir}  L. E. Ib\'{a}\~nez and G. G. Ross, 
{\it Phys. Lett. }{\bf B332 }(1994) 100;
P. Bin\'{e}truy and P. Ramond, {\it Phys.\ Lett.\ }{\bf %
B350 }(1995) 49;
V. Jain and R. Shrock, {\it Phys.\ Lett. }{\bf B352 }%
(1995) 83;
E. Dudas, S. Pokorski and C. A. Savoy, hep-ph/9504292, 
Phys. Lett. {\bf B356} (1995) 45;
Y. Nir, hep-ph/9504312, Phys. Lett. {\bf B354} (1995) 107.

\bibitem{dps}
E. Dudas, S. Pokorski and C. A. Savoy, 
{\it Phys.\ Lett. }{\bf B369 } (1996) 255;
E. Dudas, C. Grojean, S. Pokorski and C. A. Savoy, 
{\it Nucl. Phys.} {\bf B481} (1996) 85.

\bibitem{dd}
P. Bin\'{e}truy and E. Dudas, {\it Phys. Lett. }{\bf B389} (1996) 503;
G. Dvali, A. Pomarol, {\it Phys. Rev. Lett. }{\bf 77} (1996) 3728.

\bibitem{nel}
S. Ambrosiano, A. E. Nelson, {\it Phys. Lett. }{\bf B411} (1997) 283;
A. E. Nelson, D. Wright, {\it Phys. Rev.}{\bf D56 }(1997) 1598

\bibitem{jack}
I. Jack, D. R. T. Jones, S. P. Martin, M. T. Vaughn and Y. Yamada
{\it Phys. Rev.} {\bf D50} (1994) 5481.
\bibitem{poko}
M. Olechowski, S. Pokorski, {\it Phys.\ Lett.\ }{\bf B344 }(1995) 201; 
N. Polonski, A. Pomarol, {\it Phys. Rev. Lett. }{\bf 73} (1994) 2292;
D. Matalliotakis, H. P. Nilles, {\it Nucl.\ Phys.\ }{\bf B435 }(1995) 115.
P. H. Chankowski, J. Ellis, S. Pokorski, 
{\it Phys.\ Lett.\ }{\bf B423 }(1998), 327;
G. L. Kane, S. F. King, 

\bibitem{aqi} 
 L. E. Ib\'a\~{n}ez, C. Munoz, S. Rigolin, {\it Nucl.\ Phys.\ }{\bf B553}
(1999) 43; 
L.E. Ib\'a\~{n}ez, R. Rabadan , A.M. Uranga, {\it Nucl.\ Phys.\ }{\bf B542}
(1999) 112. 

\bibitem{dws}  
M. Dine, N. Seiberg and E. Witten, {\it Nucl.\ Phys.\ }%
{\bf B289 }(1987) 585;
J. Atick, L. Dixon and A. Sen, {\it Nucl.\ Phys. }{\bf B292 }(1987) 109;
M. Dine, I. Ichinose and N. Seiberg, {\it Nucl.\ Phys. }{\bf B293 }(1987)
253.
 
\bibitem{mass} 
F. Gabbiani, E. Gabrielli, A. Masiero, L. Silvestrini,
{\it Nucl. Phys.} {\bf B447} (1996) 321.

\bibitem{abe}
J.A. Casas, A. Lleyda, C. Munoz, {\it Phys. Lett. }{\bf B380} (1996) 59.
S.A. Abel, C.A. Savoy, {\it Nucl. Phys.} {\bf B532} (1998) 3.
S.A. Abel, C.A. Savoy, {\it Phys. Lett. }{\bf B444} (1998) 119.

\bibitem{alda} 
G. Aldazabal, L.E. Ibanez, F. Quevedo, {\it JHEP} {\bf 0001:031} (2000).

\end{thebibliography}
\end{document}